\documentclass[letterpaper,10pt]{article}
\usepackage{verbatim}
\usepackage{dsfont}
\usepackage{theorem}
\usepackage[numbers]{natbib}
\usepackage{setspace}
\usepackage{amsmath}
\def\T{{\hbox{\scriptsize{\rm T}}}}

\def\epsilon{\varepsilon}

\def\9dots{\vdots\,\vdots\,\vdots}
\def\phi{\varphi}

\def\th{{\hbox{\scriptsize{\rm th}}}}

\theorembodyfont{\upshape}
\newtheorem{theorem}{Theorem}
\newtheorem{lemma}[theorem]{Lemma}
\newtheorem{corollary}[theorem]{Corollary}

\newtheorem{observe}[theorem]{Observation}
\newtheorem{remark1}[theorem]{Remark}

\newenvironment{remark}{\begin{remark1} \rm}{\end{remark1}}

\usepackage{latexsym} \def\qed{\hfill $\Box$}
\newenvironment{proof}{\noindent {\bf Proof.}}{\qed\bigskip}

\title{Recurrence Relations and Fast Algorithms}

\author{Mark Tygert}

\begin{document}

\maketitle

\begin{abstract}
We construct fast algorithms for evaluating transforms
associated with families of functions which satisfy recurrence relations.
These include algorithms both for computing the coefficients
in linear combinations of the functions,
given the values of these linear combinations at certain points,
and, vice versa, for evaluating such linear combinations at those points,
given the coefficients in the linear combinations;
such procedures are also known as analysis and synthesis
of series of certain special functions.
The algorithms of the present paper are efficient in the sense that
their computational costs are proportional
to $n \, (\ln n) \, (\ln(1/\epsilon))^3$,
where $n$ is the amount of input and output data,
and $\epsilon$ is the precision of computations.
Stated somewhat more precisely,
we find a positive real number $C$ such that,
for any positive integer $n \ge 10$
and positive real number $\epsilon \le 1/10$,
the algorithms require
at most $C \, n \, (\ln n) \, (\ln(1/\epsilon))^3$ floating-point operations
and words of memory to evaluate
at $n$ appropriately chosen points
any linear combination of $n$ special functions,
given the coefficients in the linear combination,
where $\epsilon$ is the precision of computations.
\end{abstract}

\section{Introduction}

Over the past several decades,
the Fast Fourier Transform (FFT) and its variants
(see, for example,~\cite{press-teukolsky-vetterling-flannery})
have had an enormous impact across the sciences.
The FFT is an efficient algorithm for computing,
for any positive integer $n$ and complex numbers
$\beta_1$,~$\beta_2$, \dots, $\beta_{n-1}$,~$\beta_n$,
the complex numbers
$\alpha_1$,~$\alpha_2$, \dots, $\alpha_{n-1}$,~$\alpha_n$ defined by
\begin{equation}
\label{transform}
\alpha_j = \sum_{k=1}^n \beta_k \, f_k(x_j)
\end{equation}
for $j = 1$,~$2$, \dots, $n-1$,~$n$, where
$f_1$,~$f_2$, \dots, $f_{n-1}$,~$f_n$ are the functions defined on $[-1,1]$ by
\begin{equation}
\label{exponentials}
f_k(x) = \exp\left( \frac{\pi \, i \, (2k-n) \, x}{2} \right)
\end{equation}
for $k = 1$,~$2$, \dots, $n-1$,~$n$,
and $x_1$,~$x_2$, \dots, $x_{n-1}$,~$x_n$ are the real numbers defined by
\begin{equation}
\label{equispaced}
x_k = \frac{2k-n}{n}
\end{equation}
for $k = 1$,~$2$, \dots, $n-1$,~$n$.
The FFT is efficient in the sense that
there exists a reasonably small positive real number $C$ such that,
for any positive integer $n \ge 10$,
the FFT requires at most $C \, n \, \ln n$ floating-point operations
and words of memory to compute
$\alpha_1$,~$\alpha_2$, \dots, $\alpha_{n-1}$,~$\alpha_n$ in~(\ref{transform})
from $\beta_1$,~$\beta_2$, \dots, $\beta_{n-1}$,~$\beta_n$.
In contrast, evaluating the sum in~(\ref{transform}) separately
for every $j = 1$,~$2$, \dots, $n-1$,~$n$ costs at least $n^2$ operations
in total.

The present paper introduces similarly efficient algorithms
for computing
$\alpha_1$,~$\alpha_2$, \dots, $\alpha_{n-1}$,~$\alpha_n$ in~(\ref{transform})
from $\beta_1$,~$\beta_2$, \dots, $\beta_{n-1}$,~$\beta_n$,
and (when appropriate) for the inverse procedure of computing
$\beta_1$,~$\beta_2$, \dots, $\beta_{n-1}$,~$\beta_n$
from $\alpha_1$,~$\alpha_2$, \dots, $\alpha_{n-1}$,~$\alpha_n$,
for more general collections of functions $f_1$,~$f_2$, \dots, $f_{n-1}$,~$f_n$
and points $x_1$,~$x_2$, \dots, $x_{n-1}$,~$x_n$
than those defined in~(\ref{exponentials}) and~(\ref{equispaced}).
Specifically, the present paper constructs algorithms for classes of functions
satisfying recurrence relations.
The present paper describes in detail a few representative examples
of such classes of functions, namely weighted orthonormal polynomials
and Bessel functions of varying orders.
These collections of functions satisfy recurrence relations of the form
\begin{equation}
\label{general_recurrence}
g(x) \, f_k(x) = c_{k-1} \, f_{k-1}(x) + d_k \, f_k(x) + c_k \, f_{k+1}(x)
\end{equation}
for all $x$ in the domain,
where $c_{k-1}$,~$c_k$, and~$d_k$ are real numbers
and either $g(x) = x$ or $g(x) = \frac{1}{x}$;
$c_k$,~$d_k$, and~$g$ vary with the collection of functions
under consideration.

The algorithms of the present paper all rely on the following two observations:
\begin{enumerate}
\item The solutions to the recurrence relation~(\ref{general_recurrence})
are the eigenvectors corresponding to eigenvalues $g(x)$
of certain tridiagonal real self-adjoint matrices.
\item There exist fast algorithms for determining and applying matrices
whose columns are normalized eigenvectors
of a tridiagonal real self-adjoint matrix,
and for applying the adjoints of these matrices of eigenvectors.
\end{enumerate}
The first observation has been well known to numerical analysts
at least since the seminal~\cite{golub-welsch} appeared;
the second observation has been reasonably well known to numerical analysts
since the appearance of the celebrated~\cite{gu-eisenstat95}.
However, the combination seems to be new.

The methods described in the present paper should lead
to fairly efficient codes for computing a variety
of what are known as (pseudo)spectral transforms.
In particular, we can use the methods
to construct fast algorithms for calculations involving spherical harmonics
(see Remark~\ref{spherical_harmonics} below).

We refer the reader to~\cite{rokhlin-tygert} and its compilation of references
for prior work on related fast algorithms,
as well as to~\cite{jakob-chien-alpert} for an alternative approach
that is suitable for certain applications,
and to~\cite{nabors-korsmeyer-leighton-white} for its refined accounting
of computational costs.
The present paper introduces techniques that are substantially
more efficient than the extremely similar ones
for which~\cite{rokhlin-tygert} reports on far-from-optimal implementations.
We intend to report separately on carefully optimized implementations
of the techniques described in the present paper,
based in part upon the approach introduced in~\cite{martinsson-rokhlin}.
We gave a preliminary version of the present paper in~\cite{tygert_rec}.

The present paper has the following structure:
Subsection~\ref{divide-and-conquer} summarizes properties of fast algorithms
for spectral representations of tridiagonal real self-adjoint matrices,
Subsection~\ref{ortho_polys_sub} reiterates facts
having to do with recurrence relations for orthonormal polynomials,
Subsection~\ref{Bessel_funcs_sub} reiterates facts
having to do with recurrence relations for Bessel functions,
and Section~\ref{fast_algorithms} employs the subsections
of Section~\ref{prelims} to construct fast algorithms
for various purposes.

\section{Preliminaries}
\label{prelims}

This section summarizes certain widely known facts from numerical
and mathematical analysis, used in Section~\ref{fast_algorithms}.

\subsection{Divide-and-conquer spectral methods}
\label{divide-and-conquer}

This subsection summarizes properties of fast algorithms
introduced in~\cite{gu-eisenstat94} and~\cite{gu-eisenstat95}
for spectral representations of tridiagonal real self-adjoint matrices.
Specifically, there exists an algorithm such that,
for any tridiagonal real self-adjoint matrix $T$,
(firstly) the algorithm computes the eigenvalues of $T$,
(secondly) the algorithm computes any eigenvector of $T$,
(thirdly) the algorithm applies a square matrix $U$
consisting of normalized eigenvectors of $T$
to any arbitrary column vector,
and (fourthly) the algorithm applies $U^\T$ to any arbitrary column vector,
all using a number of operations and words of memory
proportional to $n \, (\ln n) \, (\ln(1/\epsilon))^3$,
where $n$ is the positive integer for which $T$ and $U$ are $n \times n$,
and $\epsilon$ is the precision of computations.
The following is a more precise formulation.

For any positive integer $n$, self-adjoint $n \times n$ matrix $T$,
and real $n \times 1$ column vector $v$,
we define $\|T\|$ to be the largest of the absolute values
of the eigenvalues of $T$, 
$\delta_T$ to be the minimum value of the distance $|\lambda-\mu|$
between any two distinct eigenvalues $\lambda$ and $\mu$ of $T$,
and
\begin{equation}
\| v \| = \sqrt{ \sum_{k=1}^n (v_k)^2 },
\end{equation}
where $v_1$,~$v_2$, \dots, $v_{n-1}$,~$v_n$ are the entries of $v$;
we say that $v$ is {\it normalized} to mean that $\| v \| = 1$.
As originated in~\cite{gu-eisenstat95},
there exist an algorithm and a positive real number $C$ such that,
for any positive real number $\epsilon \le 1/10$, positive integer $n \ge 10$,
tridiagonal real self-adjoint $n \times n$ matrix $T$
with $n$ distinct eigenvalues,
real unitary matrix $U$ whose columns are $n$ normalized eigenvectors
of $T$, and real $n \times 1$ column vector $v$,
\begin{enumerate}
\item[1.] the algorithm computes to absolute precision
$\|T\| \, \epsilon$
the $n$ eigenvalues of $T$,
using at most
\begin{equation}
C \, n \, (\ln n) \, (\ln(1/\epsilon))^3
\end{equation}
floating-point operations and words of memory,
\item[2.] the algorithm computes to absolute precision
$\|T\| \, \| v \| \, \epsilon / \delta_T$
the $n$ entries of the matrix-vector product $U \, v$,
using at most
\begin{equation}
C \, n \, (\ln n) \, (\ln(1/\epsilon))^3
\end{equation}
operations and words of memory,
\item[3.] the algorithm computes to absolute precision
$\|T\| \, \| v \| \, \epsilon / \delta_T$
the $n$ entries of the matrix-vector product $U^\T \, v$,
using at most
\begin{equation}
C \, n \, (\ln n) \, (\ln(1/\epsilon))^3
\end{equation}
operations and words of memory, and,
\item[4.] after the algorithm performs some precomputations
which are particular to $T$
at a cost of at most
\begin{equation}
C \, n \, (\ln n) \, (\ln(1/\epsilon))^3
\end{equation}
operations and words of memory,
the algorithm computes to absolute precision
$\|T\| \, \epsilon / \delta_T$
the $k \, n$ entries of any $k$ normalized eigenvectors of $T$,
using at most
\begin{equation}
C \, k \, n \, (\ln(1/\epsilon))^2
\end{equation}
operations and words of memory, for any positive integer $k$.
\end{enumerate}

\begin{remark}
We omitted distracting factors of very small powers of $n$
in the precisions mentioned in the present subsection.
Also, the bounds on the number of operations and words of memory
are extremely conservative; in actual implementations the running-times
of the algorithm appear to scale much better
with respect to the precision $\epsilon$.
\end{remark}

\begin{remark}
In the second item of the present subsection,
the algorithm in fact requires at most
\begin{equation}
C \, k \, n \, (\ln n) \, (\ln(1/\epsilon))^2
\end{equation}
operations and words of memory
to compute the matrix-vector products
$U \, v^1$, $U \, v^2$, \dots, $U \, v^{k-1}$, $U \, v^k$,
for any positive integer $k$, and real $n \times 1$ column vectors
$v^1$, $v^2$, \dots, $v^{k-1}$, $v^k$,
after the algorithm performs some precomputations
which are particular to $T$ at a cost of at most
\begin{equation}
C \, n \, (\ln n) \, (\ln(1/\epsilon))^3
\end{equation}
operations and words of memory.
Moreover, we can improve the precisions to which the algorithm calculates
$U \, v^1$, $U \, v^2$, \dots, $U \, v^{k-1}$, $U \, v^k$,
by performing more expensive precomputations
(using higher-precision floating-point arithmetic
or precomputation algorithms whose costs are not proportional
to $n \, \ln n$, for example).
Similar considerations apply to the third item of the present subsection.
\end{remark}

\begin{remark}
There exist similar algorithms
when the eigenvalues of $T$ are not all distinct.
\end{remark}

\subsection{Orthonormal polynomials}
\label{ortho_polys_sub}

This subsection discusses several classical facts concerning orthonormal
polynomials. All of these facts follow trivially from results contained,
for example, in~\cite{szego}.

Lemmas~\ref{diagonalization}, \ref{principal_lemma},
and~\ref{ortho_poly_diag}, which formulate certain simple consequences
of Theorems~\ref{rec_relation} and~\ref{distinct_theorem},
are the principal tools used
in Subsections~\ref{nodes_and_weights} and~\ref{ortho_poly_decrec}.
Lemmas~\ref{Jacobi_rec} and~\ref{Jacobi_diagonals} provide
the results of some calculations
for what are known as normalized Jacobi polynomials,
a classical example of a family of orthonormal polynomials;
the results of analogous calculations
for some other classical families of polynomials are similar
and therefore have been omitted.
The remaining lemmas in the present subsection,
Lemmas~\ref{explicit} and~\ref{Christo},
deal with certain conditioning issues surrounding the algorithms
in Subsections~\ref{nodes_and_weights} and~\ref{ortho_poly_decrec}
(see Remark~\ref{conditioning}).
The remaining theorem in the present subsection, Theorem~\ref{quadratures},
describes what are known as Gauss-Jacobi quadrature formulae.

In the present subsection, we index vectors and matrices starting
at entry $0$.

We say that {\it $a$ is an extended real number} to mean that
$a$ is a real number, $a = +\infty$, or $a = -\infty$.
For any real number $a$, we define the intervals $[a,\infty] = [a,\infty)$
and $[-\infty,a] = (-\infty,a]$;
we define $[-\infty,\infty] = (-\infty,\infty)$.

For any extended real numbers $a$ and $b$ with $a < b$
and nonnegative integer $n$,
we say that {\it $p_0$,~$p_1$, \dots, $p_{n-1}$, $p_n$
are orthonormal polynomials on $[a,b]$ for a weight $w$}
to mean that
$w$ is a real-valued nonnegative integrable function on $[a,b]$,
$p_k$ is a polynomial of degree $k$,
the coefficients of $x^0$,~$x^1$, \dots, $x^{k-1}$,~$x^k$ in $p_k(x)$ are real,
and the coefficient of $x^k$ in $p_k(x)$ is positive
for $k = 0$,~$1$, \dots, $n-1$,~$n$, and
\begin{equation}
\label{orthonormality}
\int_a^b dx \; w(x) \; p_j(x) \; p_k(x) = \left\{
\begin{array}{ll}
1, & j = k \\
0, & j \ne k
\end{array}
\right.
\end{equation}
for $j,k = 0$,~$1$, \dots, $n-1$,~$n$.

The following theorem states that a system of orthonormal polynomials
satisfies a certain three-term recurrence relation.

\begin{theorem}
\label{rec_relation}
Suppose that $a$ and $b$ are extended real numbers with $a < b$,
$n$ is a positive integer,
and $p_0$,~$p_1$, \dots, $p_{n-1}$,~$p_n$ are orthonormal polynomials
on $[a,b]$.

Then, there exist real numbers $c_0$,~$c_1$, \dots, $c_{n-2}$,~$c_{n-1}$
and $d_0$,~$d_1$, \dots, $d_{n-2}$,~$d_{n-1}$ such that
\begin{equation}
\label{recurrence0}
x \, p_0(x) = d_0 \, p_0(x) + c_0 \, p_1(x)
\end{equation}
for any $x \in [a,b]$, and
\begin{equation}
\label{recurrence}
x \, p_k(x) = c_{k-1} \, p_{k-1}(x) + d_k \, p_k(x) + c_k \, p_{k+1}(x)
\end{equation}
for any $x \in [a,b]$ and $k = 1$,~$2$, \dots, $n-2$,~$n-1$.
\end{theorem}

\begin{proof}
Theorem~3.2.1 in~\cite{szego} provides an equivalent formulation
of the present theorem.
\end{proof}

\begin{remark}
In fact, $c_k > 0$ for $k = 0$,~$1$, \dots, $n-2$,~$n-1$,
in~(\ref{recurrence0}) and~(\ref{recurrence}).
\end{remark}

The following lemma provides expressions
for $c_0$,~$c_1$, \dots, $c_{n-2}$,~$c_{n-1}$
and $d_0$,~$d_1$, \dots, $d_{n-2}$,~$d_{n-1}$
from~(\ref{recurrence0}) and~(\ref{recurrence})
for what are known as normalized Jacobi polynomials.

\begin{lemma}
\label{Jacobi_rec}
Suppose that $a = -1$, $b = 1$, $\alpha$ and $\beta$ are real numbers
with $\alpha > -1$ and $\beta > -1$, $n$ is a positive integer,
and $p_0$,~$p_1$, \dots, $p_{n-1}$,~$p_n$ are the orthonormal polynomials
on $[a,b]$ for the weight $w$ defined by
\begin{equation}
w(x) = (1-x)^\alpha \, (1+x)^\beta.
\end{equation}

Then,
\begin{equation}
\label{Jacobi_upper_diagonal}
c_k = \sqrt{
 \frac{4(k+1)(k+\alpha+1)(k+\beta+1)(k+\alpha+\beta+1)}{(2k+\alpha+\beta+1)
                            \, (2k+\alpha+\beta+2)^2 \, (2k+\alpha+\beta+3)}
}
\end{equation}
and
\begin{equation}
\label{Jacobi_diagonal}
d_k = \frac{\beta^2-\alpha^2}{(2k+\alpha+\beta)(2k+\alpha+\beta+2)}
\end{equation}
for $k = 0$,~$1$, \dots, $n-2$,~$n-1$,
where $c_0$,~$c_1$, \dots, $c_{n-2}$,~$c_{n-1}$
and $d_0$,~$d_1$, \dots, $d_{n-2}$,~$d_{n-1}$
are from~(\ref{recurrence0}) and~(\ref{recurrence}).
\end{lemma}

\begin{proof}
Formulae~4.5.1 and~4.3.4 in~\cite{szego} together provide
an equivalent formulation of the present lemma.
\end{proof}

The following theorem states that the polynomial of degree $n$
in a system of orthonormal polynomials on $[a,b]$ has $n$ distinct zeros
in $[a,b]$.

\begin{theorem}
\label{distinct_theorem}
Suppose that $a$ and $b$ are extended real numbers with $a < b$,
$n$ is a positive integer,
and $p_0$,~$p_1$, \dots, $p_{n-1}$,~$p_n$ are orthonormal polynomials
on $[a,b]$.

Then,
there exist distinct real numbers
$x_0$,~$x_1$, \dots, $x_{n-2}$,~$x_{n-1}$ such that
$x_k \in [a,b]$ and
\begin{equation}
\label{zeros}
p_n(x_k) = 0
\end{equation}
for $k = 0$,~$1$, \dots, $n-2$,~$n-1$,
and
\begin{equation}
\label{distinctness}
x_j \ne x_k
\end{equation}
when $j \ne k$ for $j,k = 0$,~$1$, \dots, $n-2$,~$n-1$.
\end{theorem}

\begin{proof}
Theorem~3.3.1 in~\cite{szego} provides a slightly more general formulation
of the present theorem.
\end{proof}

Suppose that $a$ and $b$ are extended real numbers with $a < b$,
$n$ is a positive integer,
and $p_0$,~$p_1$, \dots, $p_{n-1}$,~$p_n$ are orthonormal polynomials
on $[a,b]$ for a weight $w$.
We define $T$ to be the tridiagonal real self-adjoint $n \times n$ matrix
with the entry
\begin{equation}
\label{tridiagonal}
T_{j,k} = \left\{
\begin{array}{ll}
c_{j-1}, & k = j-1 \\
d_j, & k = j \\
c_j, & k = j+1 \\
0, & {\rm otherwise\ (when\ } k < j-1 {\rm \ or\ } k > j+1 {\rm )}
\end{array}
\right.
\end{equation}
for $j,k = 0$,~$1$, \dots, $n-2$,~$n-1$,
where $c_0$,~$c_1$, \dots, $c_{n-2}$,~$c_{n-1}$
and $d_0$,~$d_1$, \dots, $d_{n-2}$,~$d_{n-1}$
are from~(\ref{recurrence0}) and~(\ref{recurrence}).
For $k = 0$,~$1$, \dots, $n-1$,~$n$,
we define the function $q_k$ on $[a,b]$ by
\begin{equation}
\label{weighted}
q_k(x) = \sqrt{w(x)} \; p_k(x).
\end{equation}
We define $U$ to be the real $n \times n$ matrix with the entry
\begin{equation}
\label{unitary}
U_{j,k} = \frac{q_j(x_k)}{\sqrt{\sum_{m = 0}^{n-1} \left( q_m(x_k) \right)^2}}
\end{equation}
for $j,k = 0$,~$1$, \dots, $n-2$,~$n-1$,
where $q_0$,~$q_1$, \dots, $q_{n-2}$,~$q_{n-1}$ are defined
in~(\ref{weighted}),
and $x_0$,~$x_1$, \dots, $x_{n-2}$,~$x_{n-1}$ are from~(\ref{zeros}).
We define $\Lambda$ to be the diagonal real $n \times n$ matrix with the entry
\begin{equation}
\label{diagonal}
\Lambda_{j,k} = \left\{
\begin{array}{ll}
x_j, & k = j \\
0, & k \ne j
\end{array}
\right.
\end{equation}
for $j,k = 0$,~$1$, \dots, $n-2$,~$n-1$,
where $x_0$,~$x_1$, \dots, $x_{n-2}$,~$x_{n-1}$ are from~(\ref{zeros}).
We define $S$ to be the diagonal real $n \times n$ matrix with the entry
\begin{equation}
\label{scaling}
S_{j,k} = \left\{
\begin{array}{ll}
\sqrt{\sum_{m = 0}^{n-1} \left( q_m(x_j) \right)^2}, & k = j \\
0, & k \ne j
\end{array}
\right.
\end{equation}
for $j,k = 0$,~$1$, \dots, $n-2$,~$n-1$,
where $q_0$,~$q_1$, \dots, $q_{n-2}$,~$q_{n-1}$ are defined
in~(\ref{weighted}),
and $x_0$,~$x_1$, \dots, $x_{n-2}$,~$x_{n-1}$ are from~(\ref{zeros}).
We define $e$ to be the real $n \times 1$ column vector with the entry
\begin{equation}
\label{impulse}
e_k = \left\{
\begin{array}{ll}
1, & k = 0 \\
0, & k \ne 0
\end{array}
\right.
\end{equation}
for $k = 0$,~$1$, \dots, $n-2$,~$n-1$.

The following lemma states that $U$ is a matrix of normalized eigenvectors
of the tridiagonal real self-adjoint matrix $T$,
and that $\Lambda$ is a diagonal matrix
whose diagonal entries are the eigenvalues of $T$
(which, according to~(\ref{distinctness}), are distinct).

\begin{lemma}
\label{diagonalization}
Suppose that $a$ and $b$ are extended real numbers with $a < b$,
$n$ is a positive integer,
and $p_0$,~$p_1$, \dots, $p_{n-1}$,~$p_n$ are orthonormal polynomials
on $[a,b]$ for a weight $w$.

Then,
\begin{equation}
\label{diagonal_rep}
U^\T \, T \, U = \Lambda,
\end{equation}
where $T$ is defined in~(\ref{tridiagonal}),
$U$ is defined in~(\ref{unitary}),
and $\Lambda$ is defined in~(\ref{diagonal}).
Moreover, $U$ is real and unitary.
\end{lemma}

\begin{proof}
Combining~(\ref{recurrence0}), (\ref{recurrence}), and~(\ref{zeros})
yields that
\begin{equation}
\label{eigenvectors}
T \, U = U \, \Lambda.
\end{equation}
Combining~(\ref{eigenvectors}), (\ref{unitary}),
(\ref{diagonal}), and~(\ref{distinctness})
yields that $U$ is a real matrix of normalized eigenvectors of $T$,
with distinct corresponding eigenvalues.
Therefore, since eigenvectors corresponding to distinct eigenvalues
of a real self-adjoint matrix are orthogonal, $U$ is orthogonal.
Applying $U^\T$ from the left to both sides of~(\ref{eigenvectors})
yields~(\ref{diagonal_rep}).
\end{proof}

The following lemma expresses in matrix notation
the analysis and synthesis of linear combinations
of weighted orthonormal polynomials for which
Subsection~\ref{ortho_poly_decrec} describes fast algorithms.

\begin{lemma}
\label{principal_lemma}
Suppose that $a$ and $b$ are extended real numbers with $a < b$,
$n$ is a positive integer,
$p_0$,~$p_1$, \dots, $p_{n-1}$,~$p_n$ are orthonormal polynomials on $[a,b]$
for a weight $w$, and $\alpha$ and $\beta$
are real $n \times 1$ column vectors, such that $\alpha$ has the entry
\begin{equation}
\alpha_j = \sum_{k = 0}^{n-1} \beta_k \, q_k(x_j)
\end{equation}
for $j = 0$,~$1$, \dots, $n-2$,~$n-1$,
where $q_0$,~$q_1$, \dots, $q_{n-2}$,~$q_{n-1}$ are defined
in~(\ref{weighted}),
and $x_0$,~$x_1$, \dots, $x_{n-2}$,~$x_{n-1}$ are from~(\ref{zeros}).

Then,
\begin{equation}
\label{reconstruction}
\alpha = S \, U^\T \beta
\end{equation}
and
\begin{equation}
\label{decomposition}
\beta = U \, S^{-1} \, \alpha,
\end{equation}
where $U$ is defined in~(\ref{unitary}),
$S$ is defined in~(\ref{scaling}),
and $S \, U^\T \beta$ and $U \, S^{-1} \, \alpha$
are matrix-matrix-vector products.
\end{lemma}

\begin{proof}
Combining~(\ref{unitary}) and~(\ref{scaling})
yields~(\ref{reconstruction}).
According to Lemma~\ref{diagonalization}, $U$ is real and unitary.
Therefore, applying the matrix-matrix product $U \, S^{-1}$
from the left to both sides
of~(\ref{reconstruction}) yields~(\ref{decomposition}).
\end{proof}

The following two lemmas provide alternative expressions
for the entries of $S$ defined in~(\ref{scaling}).

\begin{lemma}
\label{ortho_poly_diag}
Suppose that $a$ and $b$ are extended real numbers with $a < b$,
$n$ is a positive integer,
and $p_0$,~$p_1$, \dots, $p_{n-1}$,~$p_n$ are orthonormal polynomials
on $[a,b]$ for a weight $w$.

Then,
\begin{equation}
\label{fast_diagonal}
S_{k,k} = \frac{\sqrt{w(x_k)}}{(U^\T \, e)_k \, \sqrt{\int_a^b dx \; w(x)}}
\end{equation}
for $k = 0$,~$1$, \dots, $n-2$,~$n-1$,
where $S$ is defined in~(\ref{scaling}), $U$ is defined in~(\ref{unitary}),
$e$ is defined in~(\ref{impulse}),
$(U^\T \, e)_0$,~$(U^\T \, e)_1$, \dots,
$(U^\T \, e)_{n-2}$,~$(U^\T \, e)_{n-1}$
are the entries of the matrix-vector product $U^\T \, e$,
and $x_0$,~$x_1$, \dots, $x_{n-2}$,~$x_{n-1}$ are from~(\ref{zeros}).
\end{lemma}

\begin{proof}
Combining~(\ref{unitary}) and~(\ref{impulse}) yields that
\begin{equation}
\label{matvec_eval}
(U^\T \, e)_k
= \frac{q_0(x_k)}{\sqrt{\sum_{m=0}^{n-1} \left( q_m(x_k) \right)^2}}
\end{equation}
for $k = 0$,~$1$, \dots, $n-2$,~$n-1$.
Since the polynomial $p_0$ has degree $0$,
combining~(\ref{orthonormality}) and~(\ref{weighted}) yields that
\begin{equation}
\label{degree0}
q_0(x) = \frac{\sqrt{w(x)}}{\sqrt{\int_a^b dy \; w(y)}}
\end{equation}
for any $x \in [a,b]$.
Combining~(\ref{scaling}), (\ref{matvec_eval}), and~(\ref{degree0})
yields~(\ref{fast_diagonal}).
\end{proof}

\begin{remark}
Formula~2.6 in~\cite{golub-welsch} motivated us to employ
the equivalent~(\ref{fast_diagonal}).
\end{remark}

\begin{lemma}
\label{explicit}
Suppose that $a$ and $b$ are extended real numbers with $a < b$,
$n$ is a positive integer,
$p_0$,~$p_1$, \dots, $p_{n-1}$,~$p_n$ are orthonormal polynomials
on $[a,b]$ for a weight $w$, and $k$ is a nonnegative integer, such that
$\ln w$ is differentiable at the point $x_k$ from~(\ref{zeros}).

Then,
\begin{equation}
\label{general_alternative}
(S_{k,k})^2 = c_{n-1} \, q_{n-1}(x_k) \; \frac{d}{dx} q_n(x_k),
\end{equation}
where $S_{k,k}$ is defined in~(\ref{scaling}),
$c_{n-1}$ is from~(\ref{recurrence}),
$q_{n-1}$ and $q_n$ are defined in~(\ref{weighted}),
and $x_k$ is from~(\ref{zeros}).
\end{lemma}

\begin{proof}
Formula~3.2.4 in~\cite{szego} provides a slightly more general formulation
of the present lemma.
\end{proof}

\begin{remark}
There exist similar formulations of Lemma~\ref{explicit}
when it is not the case that $\ln w$ is differentiable at $x_k$.
\end{remark}

The following theorem describes what are known as Gauss-Jacobi quad\-rature
formulae for orthonormal polynomials.

\begin{theorem}
\label{quadratures}
Suppose that $a$ and $b$ are extended real numbers with $a < b$,
$n$ is a positive integer,
and $p_0$,~$p_1$, \dots, $p_{n-1}$,~$p_n$ are orthonormal polynomials
on $[a,b]$ for a weight $w$.

Then, there exist positive real numbers
$w_0$,~$w_1$, \dots, $w_{n-2}$,~$w_{n-1}$, called the
{\rm Christoffel numbers for $x_0$,~$x_1$, \dots, $x_{n-2}$,~$x_{n-1}$},
such that
\begin{equation}
\label{Christoffel_definition}
\int_a^b dx \; w(x) \; p(x) = \sum_{k=0}^{n-1} w_k \, p(x_k)
\end{equation}
for any polynomial $p$ of degree at most $2n-1$,
where $x_0$,~$x_1$, \dots, $x_{n-2}$,~$x_{n-1}$ are from~(\ref{zeros}).
\end{theorem}

\begin{proof}
Theorems~3.4.1 and~3.4.2 in~\cite{szego} together provide a slightly more
general formulation of the present theorem.
\end{proof}

The following lemma provides alternative expressions
for the entries of $S$ defined in~(\ref{scaling}).

\begin{lemma}
\label{Christo}
Suppose that $a$ and $b$ are extended real numbers with $a < b$,
$n$ is a positive integer,
and $p_0$,~$p_1$, \dots, $p_{n-1}$,~$p_n$ are orthonormal polynomials
on $[a,b]$ for a weight $w$.

Then,
\begin{equation}
\label{Christoffel}
(S_{k,k})^2 = \frac{w(x_k)}{w_k}
\end{equation}
for $k = 0$,~$1$, \dots, $n-2$,~$n-1$,
where $S$ is defined in~(\ref{scaling}),
and $w_0$,~$w_1$, \dots, $w_{n-2}$,~$w_{n-1}$
are the Christoffel numbers from~(\ref{Christoffel_definition})
for the corresponding points $x_0$,~$x_1$, \dots, $x_{n-2}$,~$x_{n-1}$
from~(\ref{zeros}).
Moreover, there exist extended real numbers
$y_0$,~$y_1$, \dots, $y_{n-1}$,~$y_n$ such that
$a = y_0 < y_1 < \dots < y_{n-1} < y_n = b$ and
\begin{equation}
\label{Christoffel_expression}
w_k = \int_{y_k}^{y_{k+1}} dx \, w(x)
\end{equation}
for $k = 0$,~$1$, \dots, $n-2$,~$n-1$.
\end{lemma}

\begin{proof}
Formula~3.4.8 in~\cite{szego} provides an equivalent formulation
of~(\ref{Christoffel}).
Formula~3.41.1 in~\cite{szego} provides a slightly more general formulation
of~(\ref{Christoffel_expression}).
\end{proof}

\begin{remark}
\label{conditioning}
The formulae~(\ref{scaling}), (\ref{general_alternative}),
(\ref{Christoffel}), and~(\ref{Christoffel_expression}) give some insight
into the condition number of $S$.
For instance, due to~(\ref{scaling}),
the entries of $S$ are usually not too large.
\end{remark}

The following lemma provides an alternative expression
for the entries of $S$ defined in~(\ref{scaling})
for what are known as normalized Jacobi polynomials.

\begin{lemma}
\label{Jacobi_diagonals}
Suppose that $a = -1$, $b = 1$, $\alpha$ and $\beta$ are real numbers
with $\alpha > -1$ and $\beta > -1$, $n$ is a positive integer,
and $p_0$,~$p_1$, \dots, $p_{n-1}$,~$p_n$ are the orthonormal polynomials
on $[a,b]$ for the weight $w$ defined by
\begin{equation}
w(x) = (1-x)^\alpha \, (1+x)^\beta.
\end{equation}

Then,
\begin{equation}
\label{Jacobi_diagonal_expression}
S_{k,k} = \sqrt{\frac{1-x_k^2}{2n+\alpha+\beta+1}}
       \; \left| \frac{d}{dx} q_n(x_k) \right|
\end{equation}
for $k = 0$,~$1$, \dots, $n-2$,~$n-1$,
where $S$ is defined in~(\ref{scaling}),
$x_0$,~$x_1$, \dots, $x_{n-2}$,~$x_{n-1}$ are from~(\ref{zeros}),
and $q_n$ is defined in~(\ref{weighted}).
\end{lemma}

\begin{proof}
Together with~(\ref{Christoffel}),
Formulae~15.3.1 and~4.3.4 in~\cite{szego} provide an equivalent formulation
of the present lemma.
\end{proof}

\subsection{Bessel functions}
\label{Bessel_funcs_sub}

This subsection discusses several well-known facts concerning Bessel functions.
All of these facts follow trivially from results contained, for example,
in~\cite{watson} and~\cite{parlett}.

Lemmas~\ref{bounding_lemma}, \ref{principal_Bessel_lemma},
\ref{fast_Bessel_lemma}, and~\ref{Bessel_conditioning}
are the principal tools
used in Subsections~\ref{lowest_zeros} and~\ref{subevals}.
These lemmas formulate certain simple consequences
of Theorems~\ref{Bessel_recurrence} and~\ref{Sturm}
and Corollary~\ref{simplicity},
by way of Lemmas~\ref{approx_eigenvector} and~\ref{eig_bounding_lemma}.
Lemmas~\ref{Gegenbauer_lemma} and~\ref{Gegenbauer_result_lemma}
provide closed-form results of some calculations
for what are known as spherical Bessel functions,
a family of Bessel functions frequently encountered in applications.
The remaining lemmas in the present subsection,
Lemmas~\ref{arbitrary_orders} and~\ref{diagonal_for_arbitrary},
provide closed-form results of similar calculations
for Bessel functions of arbitrary nonnegative orders
(however, the results in Lemmas~\ref{arbitrary_orders}
and~\ref{diagonal_for_arbitrary} arise
from identities that are presumably not quite as familiar).

In the present subsection, we index vectors and matrices starting
at entry $1$.

Suppose that $\nu$ is a nonnegative real number.
For any nonnegative integer $k$, we define the function $f_k$ on $(0, \infty)$
by
\begin{equation}
\label{normalized_Bessel}
f_k(x) = \frac{2^\nu \, \Gamma(\nu+1) \, \sqrt{\nu+k}}{x^\nu}
       \, J_{\nu+k}(x),
\end{equation}
where $\Gamma$ is the gamma (factorial) function
and $J_{\nu+k}$ is the Bessel function of the first kind of order $\nu+k$
(see, for example, \cite{watson}).

\begin{remark}
Formula~8 of Section~3.1 in~\cite{watson} provides a more general formulation
of the fact that
\begin{equation}
\lim_{x \to 0^+} \frac{2^\nu \, \Gamma(\nu+1)}{x^\nu} \, J_{\nu}(x) = 1,
\end{equation}
which motivated our choice of normalization in~(\ref{normalized_Bessel}).
\end{remark}

The following theorem states that $f_1$,~$f_2$,~$f_3$, \dots\
defined in~(\ref{normalized_Bessel})
satisfy a certain three-term recurrence relation.

\begin{theorem}
\label{Bessel_recurrence}
Suppose that $\nu$ is a nonnegative real number.

Then,
\begin{equation}
\label{Bessel_rec0}
\frac{1}{x} \, f_1(x)
= \frac{1}{2 \sqrt{(\nu+1)}} \, \frac{2^\nu \, \Gamma(\nu+1)}{x^\nu}
\, J_{\nu}(x)
+ \frac{1}{2 \sqrt{(\nu+1)(\nu+2)}} \, f_2(x)
\end{equation}
for any positive real number $x$, and
\begin{equation}
\label{Bessel_rec}
\frac{1}{x} \, f_k(x)
= \frac{1}{2 \sqrt{(\nu+k-1)(\nu+k)}} \, f_{k-1}(x)
+ \frac{1}{2 \sqrt{(\nu+k)(\nu+k+1)}} \, f_{k+1}(x)
\end{equation}
for any positive real number $x$ and $k = 2$,~$3$,~$4$, \dots,
where $f_1$,~$f_2$,~$f_3$, \dots\ are defined in~(\ref{normalized_Bessel}),
$\Gamma$ is the gamma (factorial) function,
and $J_{\nu}$ is the Bessel function of the first kind of order $\nu$
(see, for example, \cite{watson}).
\end{theorem}

\begin{proof}
Formula~1 of Section~3.2 in~\cite{watson} provides a somewhat more general
formulation of the present theorem.
\end{proof}

Suppose that $\nu$ is a nonnegative real number and $n$ is a positive integer.
We define $T$ to be the tridiagonal real self-adjoint $n \times n$ matrix
with the entry
\begin{equation}
\label{tridiagonal_Bessel}
T_{j,k} = \left\{
\begin{array}{ll}
\frac{1}{2 \sqrt{(\nu+j-1)(\nu+j)}}, & k = j-1 \\
\frac{1}{2 \sqrt{(\nu+j)(\nu+j+1)}}, & k = j+1 \\
0,
& {\rm otherwise\ (} k < j-1,{\rm\ } k = j, {\rm \ or\ } k > j+1 {\rm )}
\end{array}
\right.
\end{equation}
for $j,k = 1$,~$2$, \dots, $n-1$,~$n$.
For any positive real number $x$,
we define $v = v(x)$ to be the real $n \times 1$ column vector with the entry
\begin{equation}
\label{vector}
v_k = \frac{f_k(x)}{\sqrt{\sum_{m = 1}^n \left( f_m(x) \right)^2}}
\end{equation}
for $k = 1$,~$2$, \dots, $n-1$,~$n$,
where $f_1$,~$f_2$, \dots, $f_{n-1}$,~$f_n$ are defined
in~(\ref{normalized_Bessel}).
For any positive real number $x$,
we define $\delta = \delta(x)$ to be the real number
\begin{equation}
\label{remainder}
\delta = \frac{1}{2 \sqrt{(\nu+n)(\nu+n+1)}}
         \frac{\left| f_{n+1}(x) \right|}{
               \sqrt{\sum_{m = 1}^n \left( f_m(x) \right)^2}},
\end{equation}
where $f_1$,~$f_2$, \dots, $f_n$,~$f_{n+1}$ are defined
in~(\ref{normalized_Bessel}).

The following lemma states that $v$ is nearly an eigenvector
of the tridiagonal real self-adjoint matrix $T$
corresponding to an approximate eigenvalue of $\frac{1}{x}$
for any positive real number $x$ such that
$J_{\nu}(x) = 0$ and $\delta$ is small.

\begin{lemma}
\label{approx_eigenvector}
Suppose that $\nu$ is a nonnegative real number
and $n$ is a positive integer.

Then,
\begin{equation}
\label{eigenvector0}
\left| (T \, v)_n - \frac{1}{x} \, v_n \right| \le \delta
\end{equation}
and
\begin{equation}
\label{eigenvector}
(T \, v)_k = \frac{1}{x} \, v_k
\end{equation}
for $k = 1$,~$2$, \dots, $n-2$,~$n-1$ and any positive real number $x$ with
\begin{equation}
\label{Bessel_zero}
J_{\nu}(x) = 0,
\end{equation}
where $T$ is defined in~(\ref{tridiagonal_Bessel}),
$v = v(x)$ is defined in~(\ref{vector}),
$(T \, v)_1$,~$(T \, v)_2$, \dots, $(T \, v)_{n-1}$,~$(T \, v)_n$
are the entries of the matrix-vector product $T \, v$,
$\delta = \delta(x)$ is defined in~(\ref{remainder}),
and $J_\nu$ is the Bessel function of the first kind of order $\nu$
(see, for example,~\cite{watson}).
\end{lemma}

\begin{proof}
Combining~(\ref{Bessel_rec}), (\ref{Bessel_rec0}), and~(\ref{Bessel_zero})
yields~(\ref{eigenvector0}) and~(\ref{eigenvector}).
\end{proof}

\begin{remark}
\label{decaying}
It is well known that,
for any positive real number $x$, the quantity $J_{\nu+n+1}(x)$
and thence $\delta$ defined in~(\ref{remainder}) decays extremely rapidly
as $n$ increases past a band around $n = x$
of width proportional to $x^{1/3}$; see, for example,
Lemma~2.5 in~\cite{rokhlin}, Chapters~9 and~10 in~\cite{abramowitz-stegun},
or Chapter~8 in~\cite{watson}.
Therefore, $\delta$ is often small for $x$ such that
$x < n$ and $J_{\nu}(x) = 0$.
\end{remark}

The following theorem states a simple Sturm sequence property
of the eigenvalues of real self-adjoint tridiagonal matrices
whose entries on the sub- and super-diagonals are nonzero.

\begin{theorem}
\label{Sturm}
Suppose that $n$ is a positive integer,
and $T$ is a tridiagonal real self-adjoint $n \times n$ matrix,
such that all entries on the sub- and super-diagonals of $T$ are nonzero.

Then, every eigenvalue of $T$ has multiplicity $1$.
\end{theorem}

\begin{proof}
Formula~7-7-1 in~\cite{parlett} provides
an equivalent formulation of the present theorem.
\end{proof}

The following corollary of Theorem~\ref{Sturm} states
a simple Sturm sequence property
of the eigenvalues of $T$ defined in~(\ref{tridiagonal_Bessel}).

\begin{corollary}
\label{simplicity}
Suppose that $\nu$ is a nonnegative real number
and $n$ is a positive integer.

Then, every eigenvalue of $T$ defined in~(\ref{tridiagonal_Bessel})
has multiplicity $1$.
\end{corollary}

The following lemma bounds the distance between an approximate eigenvalue
and the actual eigenvalue nearest to the approximation,
as well as the discrepancy
between the corresponding normalized approximate eigenvector
and a corresponding normalized actual eigenvector.

\begin{lemma}
\label{eig_bounding_lemma}
Suppose that $\gamma$, $\lambda$, and $\mu$ are real numbers,
$n$ is a positive integer,
$T$ is a real self-adjoint $n \times n$ matrix,
and $u$ and $v$ are real $n \times 1$ column vectors, such that
$\lambda$ is the eigenvalue of $T$ nearest to $\gamma$,
$\lambda$ has multiplicity $1$,
$\mu$ is the eigenvalue of $T$ nearest but not equal to $\lambda$,
\begin{equation}
T \, u = \lambda \, u,
\end{equation}
\begin{equation}
\sum_{k=1}^n (u_k)^2 = 1,
\end{equation}
and
\begin{equation}
\sum_{k=1}^n (v_k)^2 = 1.
\end{equation}

Then,
\begin{equation}
\label{eigenvalue_bound}
|\gamma - \lambda|
\le \sqrt{ \sum_{k=1}^n \Bigl( (T \, v)_k - \gamma \, v_k \Bigr)^2 },
\end{equation}
and either (or both)
\begin{equation}
\label{eigenvector_bound1}
|v_k - u_k| \le \frac{ 2
 \; \sqrt{ \sum_{k=1}^n \Bigl( (T \, v)_k - \gamma \, v_k \Bigr)^2 } }
 {|\mu - \lambda|}
\end{equation}
for $k = 1$,~$2$, \dots, $n-1$,~$n$, or
\begin{equation}
\label{eigenvector_bound2}
|-v_k - u_k| \le \frac{ 2
 \; \sqrt{ \sum_{k=1}^n \Bigl( (T \, v)_k - \gamma \, v_k \Bigr)^2 } }
 {|\mu - \lambda|}
\end{equation}
for $k = 1$,~$2$, \dots, $n-1$,~$n$,
where $(T \, v)_1$,~$(T \, v)_2$, \dots, $(T \, v)_{n-1}$,~$(T \, v)_n$
are the entries of the matrix-vector product $T \, v$.
\end{lemma}

\begin{proof}
Formula~4-5-1 in~\cite{parlett} provides an equivalent formulation
of~(\ref{eigenvalue_bound}).

The proof of Formula~11-7-1 in~\cite{parlett},
specifically Formula~11-7-3 in~\cite{parlett}
and the comment immediately following Formula~11-7-3 in~\cite{parlett},
provides a slightly more general formulation
of the fact that~(\ref{eigenvector_bound1}) holds
for $k = 1$,~$2$, \dots, $n-1$,~$n$, or that~(\ref{eigenvector_bound2}) holds
for $k = 1$,~$2$, \dots, $n-1$,~$n$.
\end{proof}

The following lemma bounds the changes in the eigenvalues
and eigenvectors induced by using the truncated matrix $T$
defined in~(\ref{tridiagonal_Bessel})
($T$ is only $n \times n$, not infinite-dimensional).

\begin{lemma}
\label{bounding_lemma}
Suppose that $\lambda$, $\mu$, $\nu$, and $x$ are real numbers,
$n$ is a positive integer,
and $u$ is a real $n \times 1$ column vector, such that $\nu \ge 0$, $x > 0$,
(\ref{Bessel_zero}) holds,
$\lambda$ is the eigenvalue of $T$ nearest to $\frac{1}{x}$,
$\mu$ is the eigenvalue of $T$ nearest but not equal to $\lambda$,
\begin{equation}
\label{eigenvector_relation}
T \, u = \lambda \, u,
\end{equation}
and
\begin{equation}
\label{eigenvector_norm}
\sum_{k=1}^n (u_k)^2 = 1,
\end{equation}
where $T$ is defined in~(\ref{tridiagonal_Bessel}).

Then,
\begin{equation}
\label{eigenvalue_approx}
\left| \frac{1}{x} - \lambda \right| \le \delta,
\end{equation}
and either (or both)
\begin{equation}
\label{discrepancy1}
\left| v_k - u_k \right|
\le \frac{2 \delta}{|\mu-\lambda|}
\end{equation}
for $k = 1$,~$2$, \dots, $n-1$,~$n$, or
\begin{equation}
\label{discrepancy2}
\left| -v_k - u_k \right|
\le \frac{2 \delta}{|\mu-\lambda|}
\end{equation}
for $k = 1$,~$2$, \dots, $n-1$,~$n$,
where $v = v(x)$ is defined in~(\ref{vector})
and $\delta = \delta(x)$ is defined in~(\ref{remainder}).
\end{lemma}

\begin{proof}
Combining Corollary~\ref{simplicity}, (\ref{eigenvalue_bound}),
(\ref{eigenvector0}), and~(\ref{eigenvector})
yields~(\ref{eigenvalue_approx}).

Combining Corollary~\ref{simplicity}, (\ref{eigenvector_bound1}),
(\ref{eigenvector_bound2}), (\ref{eigenvector0}), and~(\ref{eigenvector})
yields that~(\ref{discrepancy1}) holds
for $k = 1$,~$2$, \dots, $n-1$,~$n$, or that~(\ref{discrepancy2}) holds
for $k = 1$,~$2$, \dots, $n-1$,~$n$.
\end{proof}

Suppose that $\nu$ is a nonnegative real number
and $n$ is a positive integer.
We define $x_1$,~$x_2$,~$x_3$, \dots\ to be all of the positive real numbers
such that
\begin{equation}
\label{Bessel_zeros}
J_{\nu}(x_k) = 0
\end{equation}
for any positive integer $k$,
ordered so that
\begin{equation}
\label{ordering}
0 < x_1 < x_2 < x_3 < \dots,
\end{equation}
where $J_{\nu}$ is the Bessel function of the first kind of order $\nu$
(see, for example,~\cite{watson}).
We define $S$ to be the diagonal real $n \times n$ matrix with the entry
\begin{equation}
\label{Bessel_diagonal}
S_{j,k} = \left\{
\begin{array}{ll}
\sqrt{\sum_{m=1}^n \left( f_m(x_j) \right)^2}, & j = k \\
0, & j \ne k
\end{array}
\right.
\end{equation}
for $j,k = 1$,~$2$, \dots, $n-1$,~$n$,
where $f_1$,~$f_2$, \dots, $f_{n-1}$,~$f_n$ are defined
in~(\ref{normalized_Bessel}),
and $x_1$,~$x_2$, \dots, $x_{n-1}$,~$x_n$ are defined
in~(\ref{Bessel_zeros}) and~(\ref{ordering}).
We define $e$ to be the real $n \times 1$ column vector with the entry
\begin{equation}
\label{Bessel_impulse}
e_k = \left\{
\begin{array}{ll}
1, & k = 1 \\
0, & k \ne 1
\end{array}
\right.
\end{equation}
for $k = 1$,~$2$, \dots, $n-1$,~$n$.

The following lemma expresses in matrix notation
the evaluations of linear combinations of Bessel functions
for which Subsection~\ref{subevals} describes fast algorithms.

\begin{lemma}
\label{principal_Bessel_lemma}
Suppose that $\nu$ is a nonnegative real number, $n$ is a positive integer,
and $\alpha$ and $\beta$ are real $n \times 1$ column vectors, such that
$\alpha$ has the entry
\begin{equation}
\alpha_j = \sum_{k=1}^n \beta_k \, f_k(x_j)
\end{equation}
for $j = 1$,~$2$, \dots, $n-1$,~$n$,
where $f_1$,~$f_2$, \dots, $f_{n-1}$,~$f_n$ are defined
in~(\ref{normalized_Bessel}),
and $x_1$,~$x_2$, \dots, $x_{n-1}$,~$x_n$ are defined
in~(\ref{Bessel_zeros}) and~(\ref{ordering}).

Then,
\begin{equation}
\label{evaluation}
|\alpha_k - (S \, U^\T \beta)_k|
\le \frac{2 \; S_{k,k} \; \delta(x_k)}{|\mu_k-\lambda_k|}
\end{equation}
for any $k = 1$,~$2$, \dots, $n-1$,~$n$ such that
\begin{equation}
\label{tedious_condition}
\frac{2 \; S_{k,k} \; \delta(x_k)}{|\mu_k-\lambda_k|}
< |f_1(x_k)|,
\end{equation}
where $\lambda_k$ is the eigenvalue of $T$ defined
in~(\ref{tridiagonal_Bessel}) nearest to $\frac{1}{x_k}$,
$\mu_k$ is the eigenvalue of $T$ nearest but not equal to $\lambda_k$,
$\delta = \delta(x_k)$ is defined in~(\ref{remainder}),
$U$ is a real $n \times n$ matrix whose $k^\th$ column
is the normalized eigenvector of $T$ corresponding
to the eigenvalue $\lambda_k$
whose first entry has the same sign as $f_1(x_k)$,
$S$ is defined in~(\ref{Bessel_diagonal}),
and $(S \, U^\T \beta)_k$ is the $k^\th$ entry
of the matrix-matrix-vector product $S \, U^\T \beta$.
\end{lemma}

\begin{proof}
Combining~(\ref{discrepancy1}), (\ref{vector}),
and~(\ref{Bessel_diagonal}) yields~(\ref{evaluation}).
\end{proof}

The following two lemmas provide alternative expressions
for the entries of $S$ defined in~(\ref{Bessel_diagonal}).

\begin{lemma}
\label{fast_Bessel_lemma}
Suppose that $\nu$ is a nonnegative real number
and $n$ is a positive integer.

Then,
\begin{equation}
\label{fast_diagonal_Bessel}
\left| S_{k,k} - \frac{f_1(x_k)}{(U^\T \, e)_k} \right|
\le \frac{2 \; S_{k,k} \; \delta(x_k)}
         {|\mu_k-\lambda_k| \, \left| (U^\T \, e)_k \right|}
\end{equation}
for any $k = 1$,~$2$, \dots, $n-1$,~$n$ such that
(\ref{tedious_condition})~holds,
where $S$ is defined in~(\ref{Bessel_diagonal}),
$f_1$ is defined in~(\ref{normalized_Bessel}),
$x_1$,~$x_2$, \dots, $x_{n-1}$,~$x_n$ are defined in~(\ref{Bessel_zeros})
and~(\ref{ordering}),
$\lambda_k$ is the eigenvalue of $T$ defined
in~(\ref{tridiagonal_Bessel}) nearest to $\frac{1}{x_k}$,
$\mu_k$ is the eigenvalue of $T$ nearest but not equal to $\lambda_k$,
$\delta = \delta(x_k)$ is defined in~(\ref{remainder}),
$U$ is a real $n \times n$ matrix whose $k^\th$ column
is the normalized eigenvector of $T$ corresponding
to the eigenvalue $\lambda_k$
whose first entry has the same sign as $f_1(x_k)$,
$e$ is defined in~(\ref{Bessel_impulse}),
and $(U^\T \, e)_k$ is the $k^\th$ entry
of the matrix-vector product $U^\T \, e$.
\end{lemma}

\begin{proof}
Combining~(\ref{discrepancy1}), (\ref{vector}),
and~(\ref{Bessel_impulse}) yields that
\begin{equation}
\label{Bessel_matvec_eval}
\left| (U^\T \, e)_k
- \frac{f_1(x_k)}{\sqrt{\sum_{m=1}^{n} \left( f_m(x_k) \right)^2}} \right|
\le \frac{2 \; \delta(x_k)}{|\mu_k-\lambda_k|}
\end{equation}
for any $k = 1$,~$2$, \dots, $n-1$,~$n$ such that
(\ref{tedious_condition})~holds.
Combining~(\ref{Bessel_diagonal}) and~(\ref{Bessel_matvec_eval})
yields~(\ref{fast_diagonal_Bessel}).
\end{proof}

\begin{lemma}
\label{Bessel_conditioning}
Suppose that $\nu$ is a nonnegative real number
and $n$ is a positive integer.

Then,
\begin{equation}
\label{alternative_f}
f_1(x_k) = -\frac{2^\nu \, \Gamma(\nu+1) \, \sqrt{\nu+1}}{(x_k)^\nu}
        \, \frac{d}{dx} J_\nu(x_k)
\end{equation}
for $k = 1$,~$2$, \dots, $n-1$,~$n$,
where $f_1$ is defined in~(\ref{normalized_Bessel}),
$x_1$,~$x_2$, \dots, $x_{n-1}$,~$x_n$ are defined in~(\ref{Bessel_zeros})
and~(\ref{ordering}),
$\Gamma$ is the gamma (factorial) function,
and $J_\nu$ is the Bessel function of the first kind of order $\nu$
(see, for example,~\cite{watson}).
\end{lemma}

\begin{proof}
Formula~4 of Section~3.2 in~\cite{watson} provides a somewhat more general
formulation of~(\ref{alternative_f}).
\end{proof}

\begin{remark}
The right hand side of~(\ref{fast_diagonal_Bessel})
involves the potentially troublesome
\begin{equation}
\label{unknown_quantity}
\frac{1}{|(U^\T \, e)_k|}.
\end{equation}
However, due to~(\ref{discrepancy1}), (\ref{vector}),
and~(\ref{Bessel_diagonal}), if
\begin{equation}
\frac{2 \; \delta(x_k)}{|\mu_k-\lambda_k|}
\end{equation}
is small,
then (\ref{unknown_quantity}) is accordingly close to
\begin{equation}
\frac{S_{k,k}}{|f_1(x_k)|},
\end{equation}
which should not be unreasonably large.
\end{remark}

\begin{remark}
Numerical experiments indicate that $|\mu_k-\lambda_k|$
in~(\ref{evaluation}) and~(\ref{fast_diagonal_Bessel})
is never exceedingly small for practical ranges of $n$;
this is probably fairly easy to prove, perhaps using the properties
of Sturm sequences. The following remark appears to be relevant.
\end{remark}

\begin{remark}
Suppose that $\nu = 0$ and $n$ is a positive integer.
For any positive real number $x$, we define $\tilde{v} = \tilde{v}(x)$ to be
the real $n \times 1$ column vector with the entry
\begin{equation}
\label{opposite_vector}
\tilde{v}_k = (-1)^k \, v_k
\end{equation}
for $k = 1$,~$2$, \dots, $n-1$,~$n$,
where $v = v(x)$ is defined in~(\ref{vector}).
Then, Formula~1 of Section~3.2 in~\cite{watson}
and Formula~2 of Section~2.1 in~\cite{watson} lead to
\begin{equation}
\left| (T \, \tilde{v})_n + \frac{1}{x} \, \tilde{v}_n \right| \le \delta
\end{equation}
in place of~(\ref{eigenvector0}),
\begin{equation}
(T \, \tilde{v})_k = -\frac{1}{x} \, \tilde{v}_k
\end{equation}
for $k = 1$,~$2$, \dots, $n-2$,~$n-1$
in place of~(\ref{eigenvector}), etc.
\end{remark}

The following lemma states a special case of the Gegenbauer addition formula
for Bessel functions.

\begin{lemma}
\label{Gegenbauer_lemma}
Suppose that $\nu = \frac{1}{2}$.

Then,
\begin{equation}
\label{Gegenbauer}
\sum_{m=0}^\infty \left( f_m(x) \right)^2 = \frac{1}{2}
\end{equation}
for any positive real number $x$,
where $f_0$,~$f_1$,~$f_2$, \dots\ are defined in~(\ref{normalized_Bessel}).
\end{lemma}

\begin{proof}
Formula~3 of Section~11.4 in~\cite{watson} provides a somewhat more general
formulation of~(\ref{Gegenbauer}).
\end{proof}

The following lemma provides alternative expressions
for the entries of $S$ defined in~(\ref{Bessel_diagonal})
for what are known as spherical Bessel functions of the first kind.

\begin{lemma}
\label{Gegenbauer_result_lemma}
Suppose that $\nu = \frac{1}{2}$, $\epsilon$ is a positive real number,
and $k$ and $n$ are positive integers, such that
\begin{equation}
\label{tail}
\sum_{m=n+1}^\infty \left( f_m(x_k) \right)^2 \le \epsilon,
\end{equation}
where $f_{n+1}$,~$f_{n+2}$,~$f_{n+3}$, \dots\ are defined
in~(\ref{normalized_Bessel}),
and $x_k$ is defined in~(\ref{Bessel_zeros}) and~(\ref{ordering}).

Then,
\begin{equation}
\label{explicit_Gegenbauer}
\left| (S_{k,k})^2 - \frac{1}{2} \right| \le \epsilon,
\end{equation}
where $S_{k,k}$ is defined in~(\ref{Bessel_diagonal}).
\end{lemma}

\begin{proof}
Combining~(\ref{Bessel_diagonal}), (\ref{Gegenbauer}), (\ref{tail}),
and~(\ref{Bessel_zeros}) yields~(\ref{explicit_Gegenbauer}).
\end{proof}

The following lemma provides a closed-form expression
for the sum in~(\ref{Gegenbauer}), for any nonnegative order $\nu$.

\begin{lemma}
\label{arbitrary_orders}
Suppose that $\nu$ is a nonnegative real number.

Then,
\begin{multline}
\label{Christoffel-Darboux_Bessel}
\sum_{m=1}^\infty (f_m(x))^2
= \frac{x^2}{2 \, (\nu+1)} \, (f_1(x))^2
+ \frac{x^2}{2} \, \left( \frac{2^\nu \, \Gamma(\nu+1)}{x^\nu} \right)^2 \,
  (J_\nu(x))^2 \\
- \frac{(2 \nu + 1) \, x}{2 \sqrt{\nu+1}} \,
  \left( \frac{2^\nu \, \Gamma(\nu+1)}{x^\nu} \right) \, J_\nu(x) \; f_1(x)
\end{multline}
for any positive real number $x$,
where $f_1$,~$f_2$,~$f_3$, \dots\ are defined in~(\ref{normalized_Bessel}),
$\Gamma$ is the gamma (factorial) function,
and $J_\nu$ is the Bessel function of the first kind of order $\nu$
(see, for example,~\cite{watson}).
\end{lemma}

\begin{proof}
Formula~57.21.1 of~\cite{hansen}
provides a more general formulation of the present lemma.
See also Formula~24 of~\cite{tygert_christ} and the surrounding discussion
for a self-contained derivation of the present lemma.
\end{proof}

The following lemma provides alternative expressions
for the entries of $S$ defined in~(\ref{Bessel_diagonal}).

\begin{lemma}
\label{diagonal_for_arbitrary}
Suppose that $\nu$ and $\epsilon$ are real numbers,
and $k$ and $n$ are positive integers, such that
$\nu \ge 0$, $\epsilon > 0$, and~(\ref{tail}) holds,
where in~(\ref{tail}), $f_{n+1}$,~$f_{n+2}$,~$f_{n+3}$, \dots\ are defined
in~(\ref{normalized_Bessel}),
and $x_k$ is defined in~(\ref{Bessel_zeros}) and~(\ref{ordering}).

Then,
\begin{equation}
\label{Bessel_alternative}
\left| (S_{k,k})^2
     - \frac{(x_k)^2}{2 \, (\nu+1)} \, \left( f_1(x_k) \right)^2 \right|
\le \epsilon,
\end{equation}
where $S_{k,k}$ is defined in~(\ref{Bessel_diagonal}),
and $f_1$ is defined in~(\ref{normalized_Bessel}).
\end{lemma}

\begin{proof}
Combining~(\ref{Bessel_diagonal}), (\ref{Christoffel-Darboux_Bessel}),
(\ref{tail}), and~(\ref{Bessel_zeros})
yields~(\ref{Bessel_alternative}).
\end{proof}

\begin{remark}
As in Remark~\ref{decaying},
it is often possible to have $\epsilon$ in~(\ref{tail}),
(\ref{explicit_Gegenbauer}), and~(\ref{Bessel_alternative}) be small
for $k$ such that $x_k < n$.
\end{remark}

\section{Fast algorithms}
\label{fast_algorithms}

This section constructs efficient algorithms
for computing the quadrature nodes and Christoffel numbers
associated with orthonormal polynomials,
for computing the zeros of Bessel functions,
for the analysis and synthesis of linear combinations
of weighted orthonormal polynomials,
and for evaluations of linear combinations of Bessel functions.
We describe the algorithms
in Subsections~\ref{nodes_and_weights} and~\ref{lowest_zeros}
solely to illustrate the generality of the techniques discussed
in the present paper;
we would expect specialized schemes
such as those in~\cite{glaser-liu-rokhlin} to outperform the algorithms
described in Subsections~\ref{nodes_and_weights} and~\ref{lowest_zeros}
in most, if not all, practical circumstances.
Each subsection in the present section relies
on both Subsection~\ref{divide-and-conquer}
and either Subsection~\ref{ortho_polys_sub}
or Subsection~\ref{Bessel_funcs_sub}.

\subsection{Quadrature nodes and Christoffel numbers
            associated with orthonormal polynomials}
\label{nodes_and_weights}

The entries of~$\Lambda$ in~(\ref{diagonal_rep}) are
the nodes $x_0$,~$x_1$, \dots, $x_{n-2}$,~$x_{n-1}$
in~(\ref{Christoffel_definition}).
We can compute rapidly the entries of $\Lambda$ in~(\ref{diagonal_rep})
using an algorithm as in the first item in Subsection~\ref{divide-and-conquer},
due to~(\ref{diagonal_rep}),
since $T$ in~(\ref{diagonal_rep}) is tridiagonal, real, and self-adjoint,
$U$ in~(\ref{diagonal_rep}) is real and unitary,
and $\Lambda$ in~(\ref{diagonal_rep}) is diagonal,
with diagonal entries that according to (\ref{distinctness}) are distinct.
For the same reason, we can apply rapidly the matrix $U^\T$
to the vector $e$ in~(\ref{fast_diagonal})
using an algorithm as in the third item in Subsection~\ref{divide-and-conquer}.
We can then compute the Christoffel numbers
$w_0$,~$w_1$, \dots, $w_{n-2}$,~$w_{n-1}$ in~(\ref{Christoffel_definition})
using~(\ref{Christoffel}) and~(\ref{fast_diagonal}).

\subsection{Zeros of Bessel functions}
\label{lowest_zeros}

We can compute rapidly the zeros $x_1$,~$x_2$, \dots, $x_{n-1}$,~$x_n$
defined in~(\ref{Bessel_zeros}) and~(\ref{ordering})
for which $\delta$ defined in~(\ref{remainder}) is sufficiently small,
using~(\ref{eigenvalue_approx}) and an algorithm as in the first item
in Subsection~\ref{divide-and-conquer},
since $T$ defined in~(\ref{tridiagonal_Bessel}) is tridiagonal, real,
and self-adjoint, and (according to Corollary~\ref{simplicity})
has $n$ distinct eigenvalues.

\subsection{Analysis and synthesis of linear combinations
            of weighted orthonormal polynomials}
\label{ortho_poly_decrec}

We can apply rapidly the matrices $U$ and $U^\T$
in~(\ref{decomposition}) and~(\ref{reconstruction})
using an algorithm as in the second and third items
in Subsection~\ref{divide-and-conquer},
due to~(\ref{diagonal_rep}),
since $T$ in~(\ref{diagonal_rep}) is tridiagonal, real, and self-adjoint,
$U$ in~(\ref{diagonal_rep}) is real and unitary,
and $\Lambda$ in~(\ref{diagonal_rep}) is diagonal,
with diagonal entries that according to (\ref{distinctness}) are distinct.
Furthermore, we can apply rapidly the remaining matrices $S$ and $S^{-1}$
in~(\ref{reconstruction}) and~(\ref{decomposition}),
since $S$ and $S^{-1}$
in~(\ref{reconstruction}) and~(\ref{decomposition}) are diagonal,
once we use~(\ref{Christoffel}) and the algorithms
from~Subsection~\ref{nodes_and_weights}
to compute the entries of~$S$ and $S^{-1}$.

\begin{remark}
\label{spherical_harmonics}
Using the algorithms described in the present subsection,
we can construct fast algorithms both for computing the coefficients
in linear combinations of spherical harmonics,
given the values of these linear combinations at certain points,
and, vice versa, for evaluating such linear combinations at those points,
given the coefficients in the linear combinations.
We can handle spherical harmonics by constructing fast algorithms
for what are known as associated Legendre functions;
see, for example,~\cite{rokhlin-tygert}.
For any nonnegative integers $l$ and $m$,
the normalized associated Legendre function
of order $m$ and degree $l$ (often denoted by $\overline{P}^m_l$)
is equal to the function $q_{l-m}$ defined in~(\ref{weighted})
for the orthonormal polynomials on $[-1,1]$
for the weight $w$ defined by
\begin{equation}
w(x) = (1-x)^m \, (1+x)^m.
\end{equation}
Thus, we could utilize the algorithms discussed in the present subsection
exactly as described,
with (\ref{Jacobi_diagonal_expression}) guaranteeing
that the condition number of $S$ is never too large
for practical ranges of $l$ and $m$.
However, we would want to take advantage of the symmetries
of associated Legendre functions,
by handling the even and odd functions separately,
using the recurrence relation associated with $x^2 \; \overline{P}^m_l(x)$
instead of the recurrence relation~(\ref{recurrence}),
which is associated with $x \; \overline{P}^m_l(x)$.
We might also want to use the Christoffel-Darboux identity
to compute interpolations to and from values
at the zeros of various polynomials,
as originated in~\cite{jakob-chien-alpert} and~\cite{yarvin-rokhlin},
and subsequently optimized and extended (see, for example,
\cite{martinsson-rokhlin} and Remarks~11 and~15 in~\cite{tygert_christ}).
\end{remark}

\subsection{Evaluations of linear combinations of Bessel functions}
\label{subevals}

We can apply rapidly the matrix $U^\T$
in~(\ref{evaluation}) using an algorithm as in the third item
in Subsection~\ref{divide-and-conquer},
since $T$ defined in~(\ref{tridiagonal_Bessel}) is tridiagonal, real,
and self-adjoint, and (according to Corollary~\ref{simplicity})
has $n$ distinct eigenvalues,
and hence $U$ in~(\ref{evaluation}) can be chosen to be real and unitary.
Furthermore, we can apply rapidly the remaining matrix $S$
in~(\ref{evaluation}),
since $S$ in~(\ref{evaluation}) is diagonal,
once we use~(\ref{fast_diagonal_Bessel}),
an algorithm as in the third section in Subsection~\ref{divide-and-conquer},
and the algorithm from Subsection~\ref{lowest_zeros}
to compute the entries of~$S$.

\section*{Acknowledgements}

We thank V. Rokhlin for his immense support and many insights;
the present paper must be regarded as a natural development of our joint work
in~\cite{rokhlin-tygert}.
We particularly appreciate his detailed editorial suggestions,
as well as being able to use his codes
in order to verify numerically most of the lemmas.
We are also very pleased to thank R. R. Coifman for his unflagging interest
and many discussions,
and grateful to F. W. J. Olver for providing the citation 
mentioned in the proof of~(\ref{Christoffel-Darboux_Bessel}).


\bibliographystyle{siam}
\bibliography{newrec}

\end{document}